\documentclass[%
reprint,
amsmath,amssymb,
aps,
prab, floatfix, nofootinbib
]{revtex4-2}

\usepackage{graphicx}
\usepackage{dcolumn}
\usepackage{bm}
\usepackage{float}
\usepackage{siunitx}
\usepackage{xcolor}
\usepackage{comment}
\usepackage{placeins}
\definecolor{light-gray}{gray}{0.8}
\definecolor{prlblue}{rgb}{0.176, 0.152, 0.57}
\usepackage[colorlinks=true, linkcolor=black, citecolor=prlblue, filecolor=black, urlcolor=prlblue]{hyperref}
\usepackage{hyperref}

\begin{document}


\title{Parametric mapping of the efficiency--instability relation\\ in plasma-wakefield accelerators}

\author{O. G. Finnerud}
\thanks{o.g.finnerud@fys.uio.no}%
\author{C. A. Lindstr{\o}m }%
\author{E. Adli}
\affiliation{
 Department of Physics, University of Oslo, 0316 Oslo, Norway\\
}%


\begin{abstract}
High efficiency is essential for plasma-wakefield accelerators to be a cost-effective alternative in high-power applications, such as a linear collider. However, in a plasma-wakefield accelerator the beam-breakup instability can be seeded by a transverse offset between the driver and trailing bunch. This instability, which rapidly increases the oscillation amplitude of the trailing bunch, grows with higher power-transfer efficiency from the driver to the trailing bunch [V.~Lebedev \textit{et al}., \href{https://doi.org/10.1103/PhysRevAccelBeams.20.121301}{Phys. Rev. Accel. Beams 21, 059901 (2018)}]. In this paper, we use particle-in-cell simulations to investigate the efficiency--instability relation that constrains the driver-to-trailing-bunch power-transfer efficiency in beam-driven plasma accelerators. We test the relation using a grid of simulations across all parameters that affect the beam-breakup instability, assuming a uniform accelerating field (optimal beam loading) and no ion motion. We find that the previously proposed efficiency--instability relation represents a lower limit on the strength of the instability for a given efficiency. For each normalized wake radius, only a certain accelerating field reaches this lowest value of the transverse instability; deviating from this point can increase the growth rate by several orders of magnitude. Lastly, we highlight how the oscillation-amplitude growth of the trailing bunch can be reduced or damped with an initial uncorrelated energy spread and the presence of ion motion.
\end{abstract}

\maketitle


\section{\label{sec:level1}Introduction}
Advanced accelerator concepts that employ plasmas as accelerating structures~\cite{veksler, Tajima, beamdriven} can sustain fields up to \SI{100}{GV/m}, three orders of magnitude larger than conventional radio-frequency (RF) accelerators~\cite{clic, ILC}. Consequently, plasma-based acceleration may be a pathway toward cost-effective compact high-energy colliders. Significant milestones such as high-efficiency acceleration~\cite{Litos2014} while preserving energy spread~\cite{efficienctpreserved}, preservation of emittance~\cite{Lindstr_m_2024}, and energy gain of over \SI{42}{GeV} in a meter-scale plasma~\cite{doubling} have been achieved using an electron beam to drive the accelerating fields in the plasma. 

To be viable for colliders, the plasma-wakefield acceleration (PWFA) scheme must ensure the accelerated electron bunch is stable enough transversely to achieve high luminosity in collisions. In the blowout regime~\cite{Rosenzweig_PRA_1991}, where a driving electron bunch with a larger density than the background plasma expels all the plasma electrons in its path and forms a focusing and accelerating ion column, transverse stability is limited by the coupling between the beam and the electrons traveling in the sheath of the blowout. High driver-to-trailing-bunch power-transfer efficiency is also required in plasma-accelerator facilities~\cite{Halfh} for them to operate cost-effectively. However, increasing the driver-to-trailing-bunch power-transfer efficiency gives rise to larger growth rates of transverse instabilities. These instabilities induce an oscillation in the trailing bunch about the plasma channel axis that rapidly increases during acceleration. Such transverse instabilities lead to emittance growth, oscillation-amplitude growth, and, if not mitigated, significant charge loss in PWFAs. 

The first type of transverse instability for the blowout regime in PWFAs was discussed by Whittum \textit{et al.}~\cite{hoseinstabilityfirst} and found to have a rapid growth rate. The instability is seeded by transverse asymmetries such as tilts in the beam. The offset beam electrons perturb the path of the plasma electrons traveling in the sheath of the plasma wake. The beam oscillates about the plasma channel axis due to the focusing forces from the exposed ions. The perturbation of the plasma wake and the oscillation of the beam slices are coupled and behave as harmonic oscillators. Each longitudinal slice drives the oscillation of the slice positioned right behind it, leading to a resonance behavior whose amplitude increases exponentially towards the tail of the beam. If not mitigated, this growth continues exponentially with the propagation distance. The instability was originally discussed for single beams in plasma (i.e., the drive beam), in which case it is known as the hose instability due to its firehose-like behavior. This instability was later found to be mitigated by the inherent energy spread developed in the driver during propagation~\cite{energyspreadhosing}, as well as the non-linear focusing fields present when the transverse beam size is large~\cite{hosingstabilitylargebeamsizes}. However, these mitigation techniques are not sufficient for the trailing bunch, as future collider applications typically require sub-percent energy spread~\cite{Halfh, colliders}, and the trailing bunch must be matched to small bunch sizes to preserve emittance~\cite{Mehrling2012,Lindstr_m_2024}. When discussing the trailing bunch, this instability is often referred to as the beam-breakup instability and is conceptually similar to that discussed in conventional RF accelerators since the 1960s~\cite{beam-breakup}. 

The strength of the instability can be characterized by the parameter~\cite{lebedev}
\begin{equation}
    \label{etat}
    \eta_{t} = -\frac{F_{t}}{F_{r}},    
\end{equation}
where $F_{t}$ is the transverse deflection force and $F_{r}$ is the ion focusing force. For small offsets, $F_{t}$ and $F_{r}$ increase linearly with offset $r$ from the axis. Hence, $\eta_{t}$ is independent of offset. Recently, Lebedev \textit{et al.}~\cite{lebedev} proposed an inherent link between the power-transfer efficiency from the driving to the trailing bunch and the strength of the beam-breakup instability, assuming no ion motion or energy spread and a uniform accelerating field experienced by the trailing bunch. The predicted $\eta_{t}$ can be expressed in terms of the efficiency $\eta_{p}$ as ~\cite{lebedev} 
\begin{equation}
    \eta_{t, \mathrm{predicted}} = \frac{\eta_{p}^{2}}{4(1-\eta_{p})},
    \label{etatetap}
\end{equation}where $\eta_{p}$ is the power-transfer efficiency from the wake to the trailing bunch. The resulting oscillation-amplitude growth of the tail particle is given as~\cite{lebedev}
\begin{equation}
\frac{A}{A_{0}} = \mathrm{exp}\left(\frac{(\mu \eta_{t})^{2}}{10 + 1.4(\mu\eta_{t})^{1.57}}\right),
\label{amplitudegrowth}
\end{equation}
where $A$ and $A_{0}$ are the oscillation amplitudes at a point along the acceleration length and the initial oscillation amplitude~\cite{aksoy}, respectively, $\mu$ is the betatron phase advance expressed in terms of the Lorentz gamma factor of the electron bunch before and after acceleration, $\gamma_{i}$ and $\gamma_{f}$, respectively, as~\cite{lebedev}
\begin{equation}
    \mu = \sqrt{2}(\sqrt{\gamma_{f}} - \sqrt{\gamma_{i}})\frac{E_{0}}{E_{z}}, 
\end{equation}
with 
\begin{equation}
    E_{0} = \frac{en_{0}}{\epsilon_0 k_{p}}
\end{equation}being the cold nonrelativistic wavebreaking field, where $k_{p}$ is the plasma wave number and $E_{z}$ is the longitudinal accelerating field. Lebedev \textit{et al}. note that the factor $(1-\eta_{p})$ in the denominator is determined by the particular dependence of the transverse wake on $\xi$, the comoving coordinate, while $\eta_{p}^{2}/4$ is universal to any structure. 

The normalized deflection force (Eq.~\ref{etatetap}) grows quadratically at lower efficiencies and even faster at higher efficiencies when the $1/(1-\eta_{p})$ term becomes significant. This suggests that the efficiencies that can be achieved in PWFAs before the strength of the deflecting wakefields in the beam-breakup (BBU) instability rises to extreme values are limited. Because of the exponential oscillation-amplitude growth at large values of $\eta_{t}$, the beam quality in transverse phase space would be expected to degrade severely at high efficiency.

However, recent work has questioned the universality of the relation in Eq.~\ref{etatetap}~\cite{baturin}. In this paper, we use the particle-in-cell (PIC) code HiPACE++~\cite{HiPACE++} to scan the entire parameter space that can affect the strength of the beam-breakup instability to probe the universality of the efficiency--instability relation.

\section{Mapping the parameter space with only three variables}
We aim to scan the entire parameter space that affects the strength of the BBU instability. The parameters chosen can be scaled to any plasma density. We turn off ion motion and suppress the energy spread by only working with optimally loaded wakefields and assuming a cold plasma. We transversely offset the trailing bunch by 0.01/$k_{p}$, sufficiently small to perturb the shape of the plasma wake only weakly. The plasma wake has a thin electron sheath~\cite{W.LU}, such that at the longitudinal position where the blowout is at its largest, the only contribution of the driver is to determine the maximum radius of the blowout $R_{b}$. This is the first parameter we consider in our parameter mapping. The second parameter to consider is the strength of the accelerating field, which is determined by the longitudinal position of the head of the trailing bunch in the plasma wake for a given $R_{b}$. The third parameter is the bunch length of the accelerated bunch. When we scan the bunch length, the charge of the trailing bunch is not a free parameter, as there is only one current profile that can load a perfectly uniform field at a given accelerating gradient and wake radius. The parameters that we include in our grid are, therefore: (1) the normalized wake radius $R_{b}k_{p}$; (2) the normalized accelerating field $E_{z}/E_{0}$; and (3) the bunch length of the trailing bunch.

\begin{figure}[h]
    \centering
    \includegraphics[width = 1\linewidth]{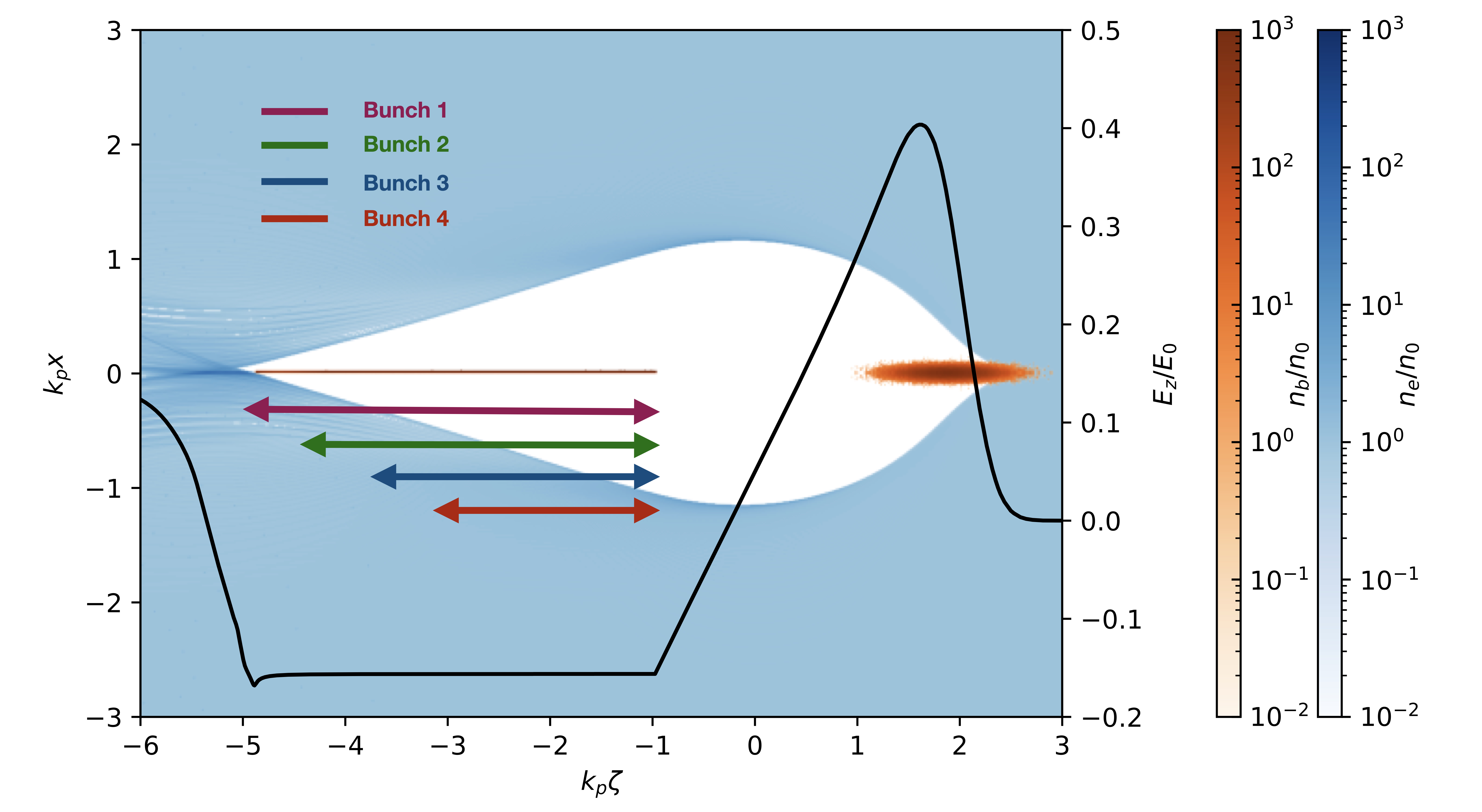}
    \caption{Plasma-electron density (blue) and beam-electron density (orange) from a particle-in-cell simulation. The solid black line shows the flattened longitudinal wakefield. Different colored lines indicate different bunch lengths that can be compared within the same simulation. For a given loading of the accelerating field, longer bunches will reach higher drive-to-trailing-bunch power-transfer efficiency. The beam is offset by 0.01/$k_{p}$. The total bunch length extends to the end of the blowout.}
    \label{fig1}
\end{figure}

\begin{figure*}[t]
    \includegraphics[width = 1\linewidth]{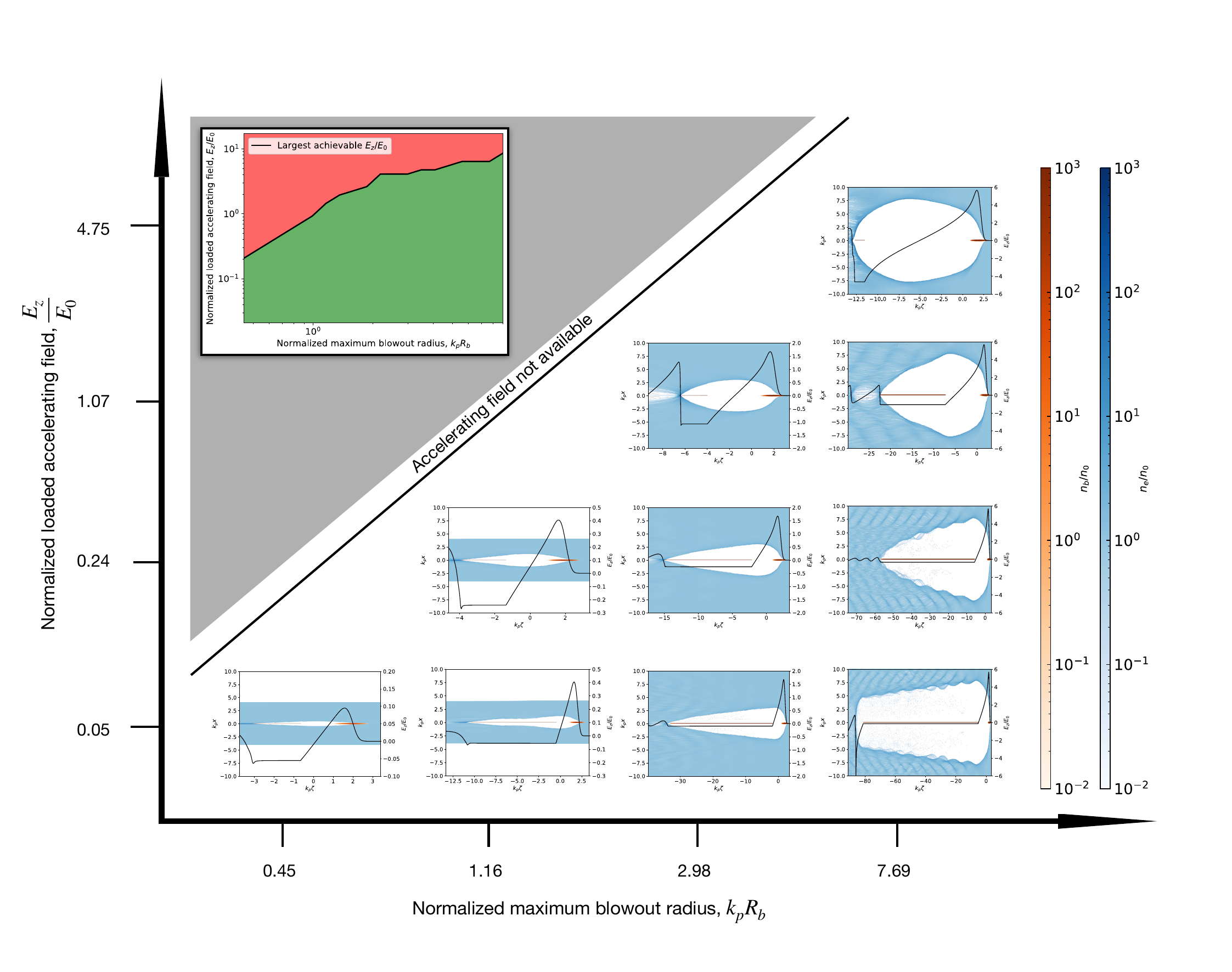}
    \caption{Grid schematic illustrating the parameter space (selected points only shown). In one dimension, we scan over the blowout wake size $R_{b}k_{p}$. The driver changes only in current along the $R_{b}k_{p}$ direction. In the other dimension, we scan over the normalized accelerating field that the trailing bunch is sampling $E_{z}/E_{0}$. Along the $E_{z}/E_{0}$ direction, the driver is identical, but the longitudinal offset of the trailing bunch changes and the beam current profile is adjusted to beam load optimally. Each simulation also provides a scan over the bunch length. The top left inset in the grey region shows the full range of achievable normalized accelerating fields for each value of $R_{b}k_{p}$ in green.} 
    \label{gridschem}
\end{figure*}

\section{Simulation procedure}
The 3D particle-in-cell (PIC) code HiPACE++ was used to simulate the plasma interaction. We employ the SALAME algorithm~\cite{SALAME} to generate the current profile that optimally beam loads the wakefield to be uniform~\cite{Tzoufras_PRL_2008}. This algorithm adjusts the weight of each longitudinal slice in the beam such that the current profile is precisely matched to the profile needed to load a uniform field.

After loading the field with SALAME, we can reduce the parameter space by one dimension, significantly speeding up the grid scan, by noticing that longitudinal slices do not affect those ahead of them. This means we can simulate only the longest bunch and subsequently extract the results for any bunch length by simply ignoring particles behind the tail location corresponding to that bunch length. Hence, we can evaluate each point in the bunch as the tail of a separate bunch, sampling the same field for each bunch-length configuration. This is indicated in Fig.~\ref{fig1}.

We evaluate the instantaneous drive-to-trailing-bunch power-transfer efficiency as 
\begin{equation}
    \eta_{p} = \frac{Q_\mathrm{trailing}E_{z}L}{\Delta U_{\mathrm{driver}}}, 
\label{eff}
\end{equation}
where $\Delta U_{\mathrm{driver}}$ is the average energy loss of the driver particles over an acceleration length $L$ times the total charge of the driver. As seen from Eq.~\ref{eff}, multiplying the integrated charge from the head of the trailing bunch to a given point along the bunch with the value of the accelerating field gives an estimate for the instantaneous efficiency $\eta_{p}$ along the beam. Only a single-step simulation is required to extract the efficiency as we only need to initialize the value of the flattened fields and the beam charges at the initial time step. Hence, we can investigate a large parameter space with a reduced computational cost as we do not require an acceleration length as a third dimension.

Figure~\ref{gridschem} shows a schematic representing the parameter scan over a range of $R_{b}k_{p}$ that covers both the non-relativistic ($R_{b}k_{p} \ll 1 $) and the ultra-relativistic ($R_{b}k_{p} \gg 1$)~\cite{W.LU} blowout regime. The normalized wake radius $R_{b}k_{p}$ was scanned over the values 0.45 to 9 with 20 values, logarithmically spaced. We choose a range of $E_{z}/E_{0}$ that contains most of the fields possible to load in the plasma wake, ranging from very long bunches where the head of the bunch is positioned further towards the front of the wake and sampling smaller accelerating fields, to ultra-short bunches positioned further towards the back of the wake and sampling larger accelerating fields. Lastly, the normalized accelerating field that the trailing bunch samples, $E_{z}/E_{0}$, was scanned over the range 0.03 to 10 with 40 values, logarithmically spaced. 

For the lower range of $E_{z}/E_{0}$, the bunch length found from SALAME required to flatten each field becomes long enough to oscillate through several plasma wavelengths—while physically possible, these solutions are unusable in practice. Therefore, we calculate only the normalized deflection force over the $z$-locations defined by the bunch electrons that satisfy the condition
\begin{equation}
    \Delta z < \mathrm{max}\left(\frac{\pi R_{b}}{2}, \frac{\lambda_{p}}{2}\right),
\end{equation}
where $\lambda_{p} = \frac{2\pi}{k_{p}}$ is the plasma wavelength. That is, at lower $R_{b}k_{p}$, we limit the $z$-locations to half a plasma wavelength, which samples the full phase in which the transverse deflecting wakefields strictly increase towards the tail of the accelerated bunch. At larger wake sizes, greater ranges of $z$-locations can sample the same field over a larger distance, hence we define a larger limiting condition constrained by the maximum wake radius $R_{b}$.

\section{Results}
\begin{figure}[t]
    \centering
    \includegraphics[width=1\linewidth]{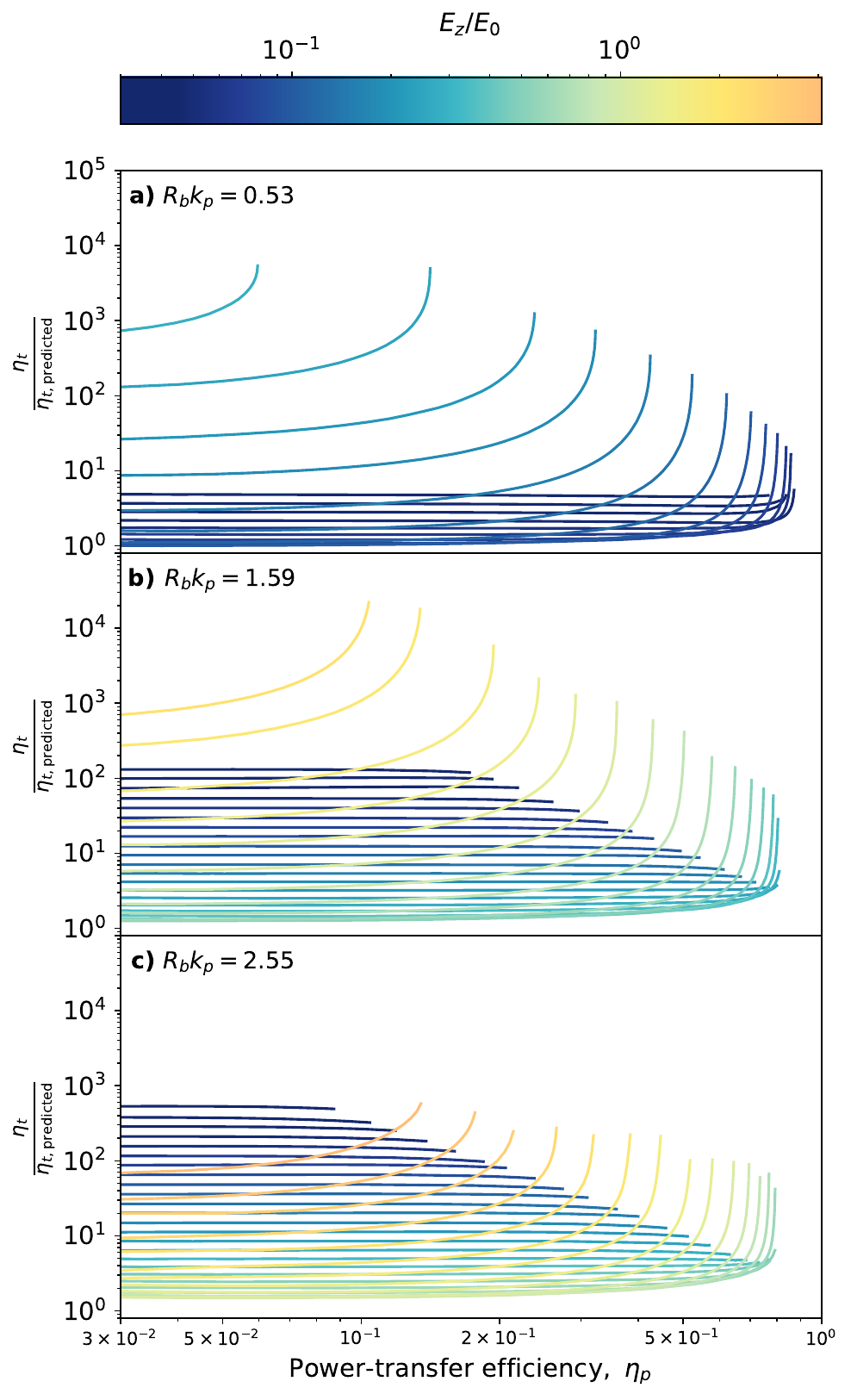}
    \caption{The strength of the instability divided by the expected analytic value from Ref.~\cite{lebedev}. The value of $\eta_{t}/\eta_{t, \mathrm{predicted}}$ 
    is shown against the drive-to-trailing-bunch instantaneous power-transfer efficiency $\eta_{p}$ for a value of the normalized wake radius $R_{b}k_{p}$ of 0.53 (a), 1.59 (b) and 2.55 (c). A blue--orange color map indicates the value of $E_{z}/E_{0}$.}
    \label{fig3}
\end{figure}

Figure~\ref{fig3} shows the correlation between $\eta_{t}/\eta_{t, \mathrm{predicted}}$ and $\eta_{p}$ for $R_{b}k_{p}$ = 0.53, $R_{b}k_{p}$ = 1.59 and $R_{b}k_{p}$ = 2.55 respectively. The normalized $\eta_{t}$ is calculated at the tail particle position of the trailing bunch. We scan over the trailing bunch from head to tail and calculate the efficiency evolution by substituting the flattened field $E_{z}$ and the charge along the bunch into Eq.~\ref{eff}. The value of $\eta_{t}/\eta_{t, \mathrm{predicted}}$ against $\eta_{p}$ is then evaluated at discrete values of $R_{b}k_{p}$. 

We can see that at each $R_{b}k_{p}$, there is a value of $E_{z}/E_{0}$ that gives the lowest value of $\eta_{t}$ possible which agrees well with the predicted value. The lowest value never goes below 1 across all parameters, but does approach it, implying that Eq.~\ref{etatetap} defines a lower bound. As the value of $E_{z}/E_{0}$ becomes larger than the optimal value, $\eta_{t}$ quickly increases by orders of magnitude. It therefore appears that this relation is a universal lower bound that can be achieved with the correct loading. Operating at the optimal point in the parameter space is crucial; otherwise, there is a significantly stronger instability. Hence, choosing the correct $E_{z}/E_{0}$ value for each $R_{b}k_{p}$ is essential.

\begin{figure}[t]
    \centering
    \includegraphics[width=1\linewidth]{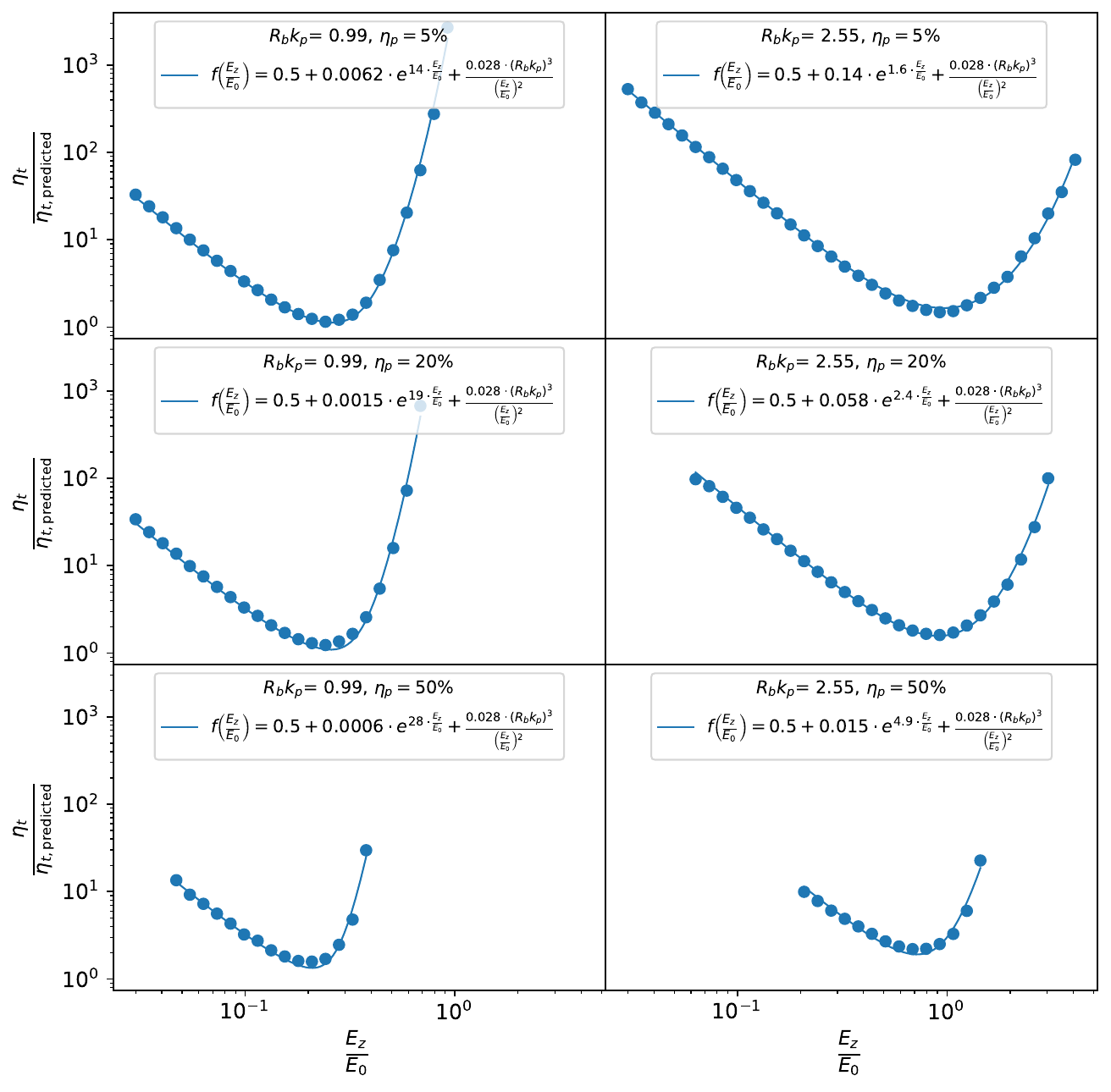}
    \caption{The evolution of $\eta_{t}/\eta_{t, \mathrm{predicted}}$ for three power-transfer efficiencies at both $R_{b}k_{p}$ = 0.99 and 2.55, as indicated in the legends. For smaller values of $E_{z}/E_{0}$ the value of $\eta_{t}/\eta_{\mathrm{t, predicted}}$ scales as $(E_{z}/E_{0})^{-2}$, whereas when $E_{z}/E_{0}$ becomes larger than the optimal value, $\eta_{t}/\eta_{t, \mathrm{predicted}}$ begins to increase exponentially. The curves represent the fit shown in the legends.}
    \label{fig4}
\end{figure} 

As the value of $E_{z}/E_{0}$ increases and approaches the optimal value, the value of $\eta_{t}/\eta_{t, \mathrm{predicted}}$ decreases at a rate that is roughly constant in log-log space. However, as the value of $E_{z}/E_{0}$ passes the optimal value and begins to increase again, the value of $\eta_{t}$ becomes larger at an increasing rate. In particular, $\eta_{t}/\eta_{t, \mathrm{predicted}}$ for the cases with fields larger than the optimal, increases exponentially when the power-transfer efficiency $\eta_{p}$ tends towards 1. This is shown in Fig.~\ref{fig4} for the normalized maximum blowout radii of $R_{b}k_{p} = 0.99$ and $R_{b}k_{p} = 2.55$ and power-transfer efficiencies of 5\%, 20\% and 50\%. We fit the $\eta_{t}/\eta_{t, \mathrm{predicted}}$ data points with values of $E_{z}/E_{0}$ less than the optimal value $E_{z, \mathrm{opt}}$, which is obtained when
$\eta_{t}/\eta_{t, \mathrm{predicted}}$ is minimized, against $E_{z}/E_{0}$ and $R_{b}k_{p}$ for a generic power-law relationship. The resulting equation 
\begin{equation}
    \eta_{t}/\eta_{t, \mathrm{predicted}}\approx \frac{0.028 \cdot (R_{b}k_{p})^3}{(\frac{E_{z}}{E_{0}})^2};\hspace{1cm} E_{z} < E_{z, \mathrm{opt}} 
    \label{evolfunc}
\end{equation}
gives very good agreement with the simulated results. 

\begin{figure}[t]
  \centering
  \includegraphics[width=0.95\linewidth]{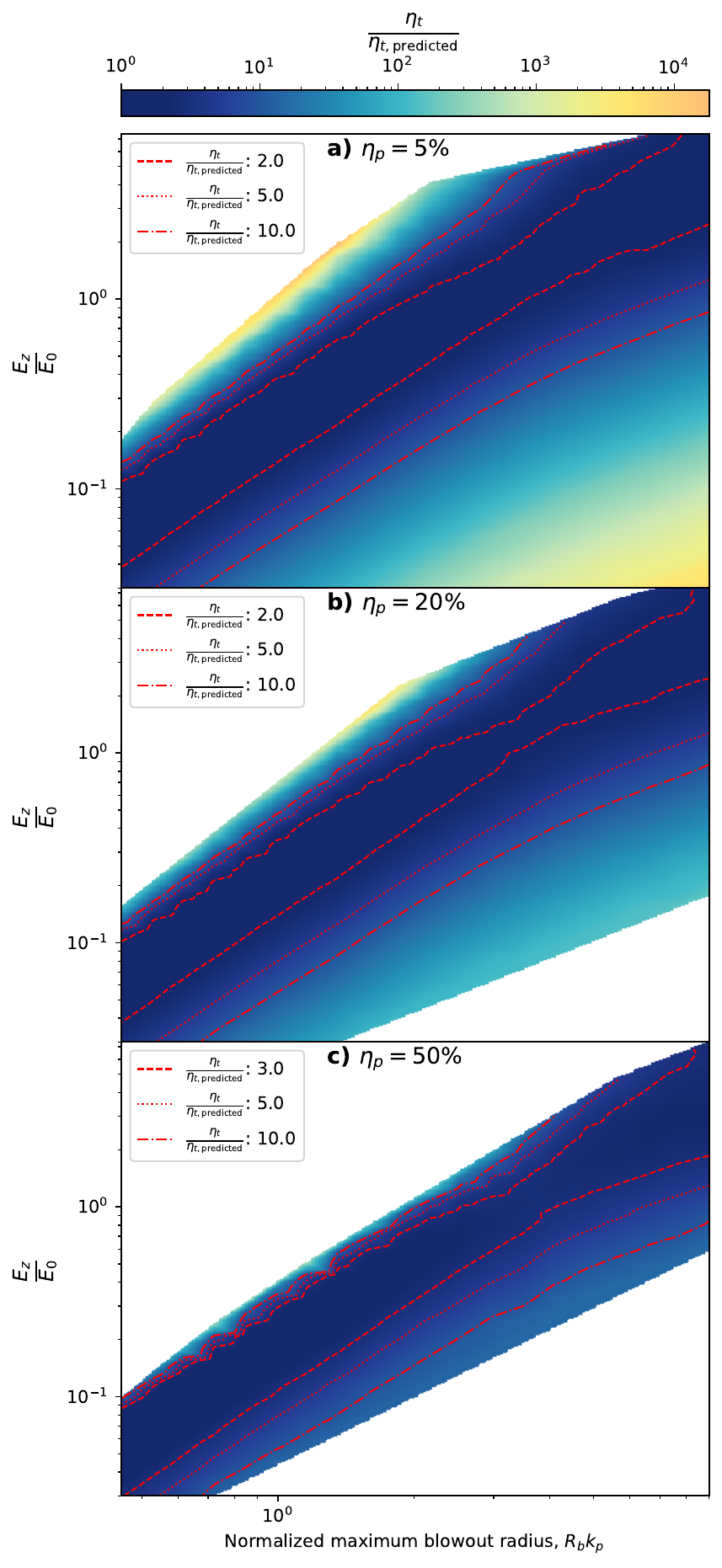}
  \caption{The strength of the instability compared to the predicted value (Eq.~\ref{etatetap}), $\eta_{t}/\eta_{t, \mathrm{predicted}}$ (blue--orange color map), is shown for all values of $R_{b}k_{p}$ and $E_{z}/E_{0}$ at (a) 5\%, (b) 20\% and (c) 50\% power-transfer efficiency $\eta_{p}$. The red contour lines enclose regions where $\eta_{t}/\eta_{t, \mathrm{predicted}}$ is constant as specified in the legends.}
  \label{fig5}
\end{figure}

At values of $E_{z}/E_{0}$ larger than optimal, the value of $\eta_{t}/\eta_{t, \mathrm{predicted}}$ increases exponentially, and the growth rate is larger for higher efficiencies. A possible explanation for this trend is that, at larger accelerating fields, the accelerated bunch is offset longitudinally further back in the blowout wake. Here, the radius of the electron sheath is rapidly decreasing and the angle at which electrons return to the axis is rapidly increasing. Therefore the coupling between the electron sheath and the offset trailing-bunch electrons will grow as the distance between them decreases, and do so at an increasing rate. However, by loading at lower accelerating fields (i.e.,~positioned further forward), the distance between the electron sheath and the offset trailing-bunch electrons will be larger and relative changes to the blowout radius throughout the bunch are small. Thus, loading at a lower than optimal field is preferable to loading at a higher field.

The entire parameter space for a given efficiency in terms of $\eta_{t}$ is shown in Fig.~\ref{fig5}, which shows complete mappings of the parameter space for three different efficiencies of 5\%, 20\%, and 50\%. The continuous map in the figure is obtained by interpolating between values of $E_{z}/E_{0}$ and $R_{b}k_{p}$ used in the simulations. There is an optimal valley in the parameter space along the range of $R_{b}k_{p}$ where the simulated $\eta_{t}$ is close to $\eta_{t, \mathrm{predicted}}$. Deviating from this region across all values of $R_{b}k_{p}$ rapidly increases $\eta_{t}/\eta_{t, \mathrm{predicted}}$, particularly while going to larger values of $E_{z}/E_{0}$. 

\begin{figure}[b]
    \includegraphics[width = 1\linewidth]{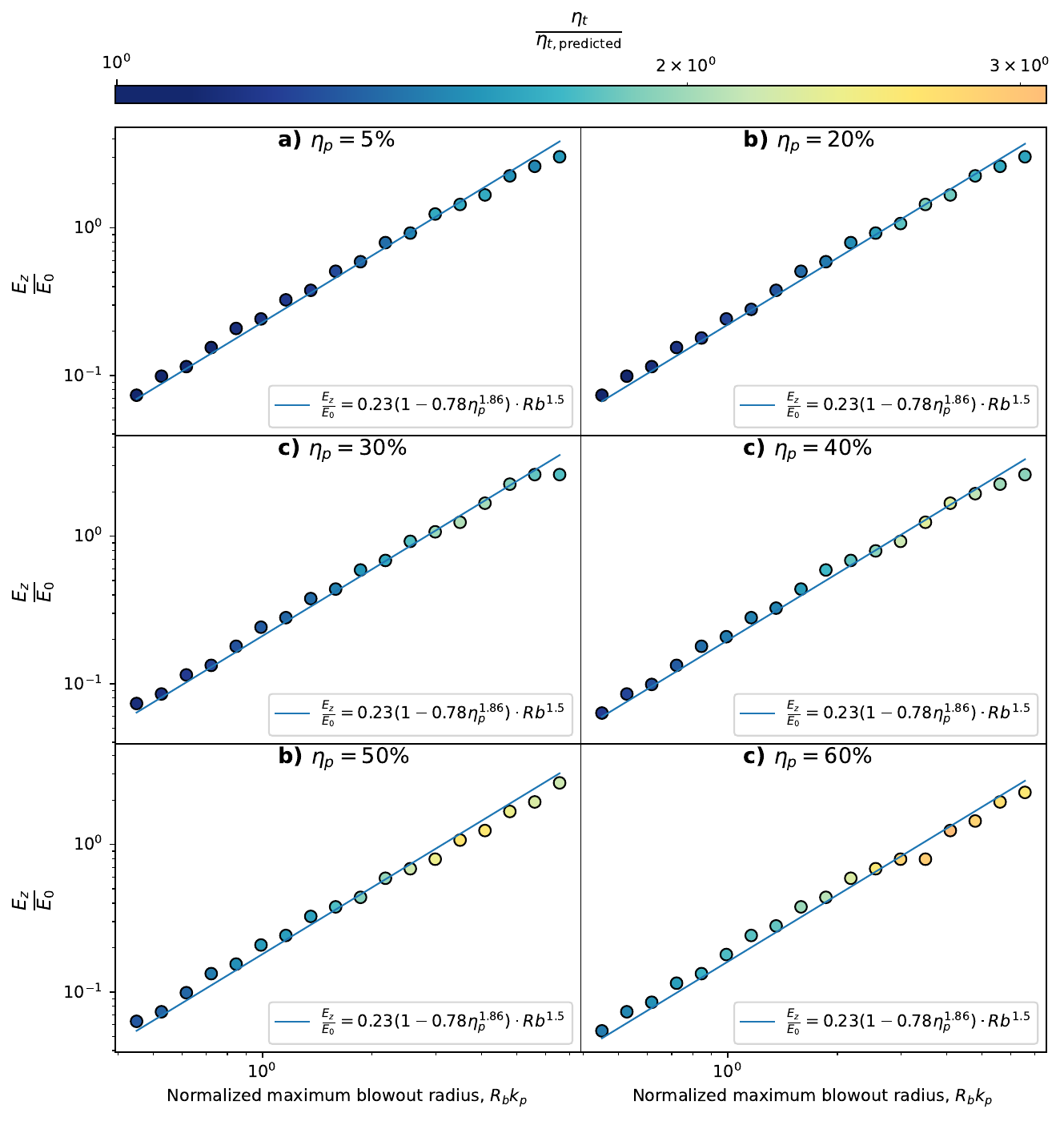}
    \caption{The optimal value of $E_{z}/E_{0}$ for each value of $R_{b}k_{p}$ that gives the lowest $\eta_{t}/\eta_{t, \mathrm{predicted}}$ attainable at power-transfer efficiencies $\eta_{p}$ of 5--60\%. The color of each point shows the value of $\eta_{t}/\eta_{t, \mathrm{predicted}}$ at that combination of $R_{b}k_{p}$ and $E_{z}/E_{0}$. The blue line indicates the best fitted line for estimating the optimal $E_{z}/E_{0}$ at each $R_{b}k_{p}$.}
    \label{fig6}
\end{figure}

Figure~\ref{fig6} shows the value of $E_{z}/E_{0}$ for which $\eta_{t}/\eta_{t, \mathrm{predicted}}$ is minimized at each $R_{b}k_{p}$ for various power-transfer efficiencies. The expected relation can be found using Eq.~\ref{evolfunc} by considering where $\eta_{t}/\eta_{t, \mathrm{predicted}} \propto (R_b k_p)^3/(E_z/E_0)^2$ intersects the value 1 (i.e., close to the minimum); solving for $\eta_{t}/\eta_{t, \mathrm{predicted}} = 1$ gives $E_z/E_0 \propto (R_b k_p)^{1.5}$. This power-law scaling is confirmed by performing a fit of $E_{z}/E_{0}$ versus $R_{b}k_{p}$ and $\eta_p$, giving
\begin{equation}
    \label{optimalEz_low}
    \frac{E_{z, \mathrm{opt}}}{E_{0}} \approx 0.23(1-0.78\eta_{p}^{1.86})(R_{b}k_{p})^{1.5}.
\end{equation}
This equation indicates the optimal accelerating field for minimizing transverse instability, and works for efficiencies up to $\sim$60\%\footnote{We exclude the two highest $R_{b}k_{p}$ values to improve the fit.}. The optimal values of $E_{z}/E_{0}$ found from simulation do not deviate by more than 10\% at low efficiency and 14\% at high efficiency from the predicted values given by Eq.~\ref{optimalEz_low}. 

The minimum attainable $\eta_{t}/\eta_{t, \mathrm{predicted}}$ (indicated with the color of each point) is less than 2 for all points at efficiencies less than 20\%. For larger efficiencies up to 60\%, all minimum attainable $\eta_{t}/\eta_{t, \mathrm{predicted}}$ are less than 4 and the larger values of $\eta_{t}/\eta_{t, \mathrm{predicted}}$ are generally reached at large values of $R_{b}k_{p}$. For smaller efficiencies and low $R_{b}k_{p}$, the optimally loaded accelerating field generally gives values of $\eta_{t}/\eta_{t, \mathrm{predicted}}$ very close to 1.

\section{Amplitude growth}
\begin{figure*}[t]  
    \includegraphics[width = 1\linewidth]{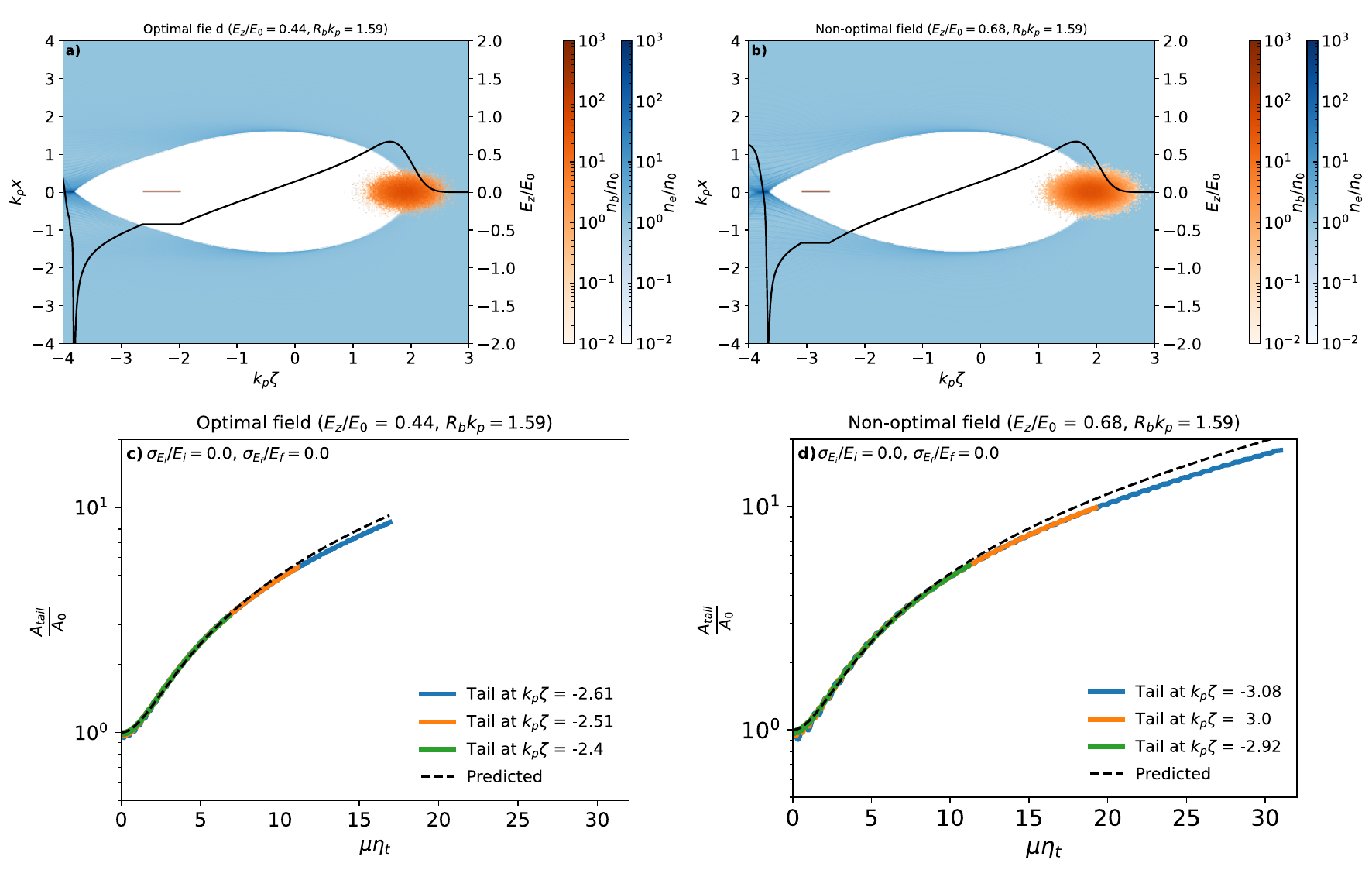}
    \caption{Initial charge density (top) for the bunches and plasma in the blowout regime immediately after the plasma wake is formed. The black line shows the accelerating field $E_{z}$. The charge densities for the lower $\eta_{t}$ configuration is shown in (a), whereas the charge density for the higher $\eta_{t}$ configuration is shown in (b). The tail oscillation-amplitude growth is shown for both the optimal-$\eta_{t}$ (c) and high-$\mathrm{\eta_{t}}$ (d) case. Three bunch lengths for each configuration are indicated: long (blue line), medium (orange line), and short (green line). In all cases, the oscillation-amplitude growth agrees reasonably well with the analytical expression from Eq.~\ref{amplitudegrowth} (dashed black line) calculated using the initial $\eta_{t}$. No energy spread or ion motion is included.}
    \label{fig:combined}
\end{figure*}

We have now established how $\eta_{t}$ depends on the parameters $\eta_{p}$, $E_{z}/E_{0}$ and $R_{b}k_{p}$. This section aims to test whether Eq.~\ref{amplitudegrowth} accurately predicts the oscillation-amplitude growth for a given $\eta_{t}$. 

Long simulations were performed (see Fig.~\ref{fig:combined}) to compare the amplitude growth between two operating points: one with a low value of $\eta_{t}/\eta_{t, \mathrm{predicted}} = 1.6$ (with field $E_{z}/E_{0} = 0.44$, close to the optimum predicted by Eq.~\ref{optimalEz_low}); and one with a higher value of $\eta_{t}/\eta_{t, \mathrm{predicted}} = 3.1$ (with field $E_{z}/E_{0} = 0.68$). In both these simulations, the normalized wake radius is $R_{b}k_{p} = 1.59$ and the efficiency is $\eta_{p} = 0.4$. The acceleration length is \SI{3.8e4}{}/$k_{p}$. The driver has normalized charge of $\tilde{Q} = 5.5$ \cite{Barov:2002bg}, rms bunch length 0.25/$k_{p}$, and a transverse rms bunch size of 0.19/$k_{p}$. The initial energy spread is set to zero and no ion motion is included. To avoid head erosion \cite{An2013} of the drive bunch, its energy is set to be very high such that its evolution is frozen along the acceleration length. Figure~\ref{fig:combined} shows that the electron bunch in the optimal-$\eta_{t}$ case (a) must be longer to achieve the same efficiency as the high-$\eta_{t}$ case (b) and is positioned further forward in the wake to load a weaker accelerating field $E_{z}$.

The oscillation-amplitude growth of the tail of the trailing bunch in each case is evaluated against the normalized quantity $\mu \eta_{t}$ [see Fig.~\ref{fig:combined}(c) and (d)], which represents the propagation distance weighted by the instability parameter $\eta_{t}$. 
The high-$\eta_{t}$ configuration (d) gives rise to higher (roughly by a factor 2) oscillation-amplitude growth compared to the optimal-$\eta_{t}$ configuration (c) at the end of the propagation distance. Note that while the latter sees more energy gain, there is less charge such that the efficiency, as defined in Eq.~\ref{eff}, is the same for both configurations; the overall energy extracted from the wake is the same in both cases. Importantly, both configurations follow the analytical expression in Eq.~\ref{amplitudegrowth} closely.

This conclusion also holds for bunches of different length. Figure~\ref{fig:combined} shows that longer bunches experience more oscillation-amplitude growth when loaded at the same accelerating field, in accordance with Eq.~\ref{amplitudegrowth}. Increasing the bunch length while preserving the perfectly flat loaded field requires adding more charge to the tail of the bunch. Hence, making the bunch longer also increases the instantaneous power-transfer efficiency $\eta_{p}$, which again increases the instability parameter $\eta_{t}$.

\section{Mitigation techniques}

The findings in this paper confirm that in the case of no energy spread or ion motion, the beam-breakup instability can be severe. This section briefly examines how the instability can be mitigated by energy spread and non-negligible ion motion.

\begin{figure*}[tb]  
    \centering
    \includegraphics[width=0.95\linewidth]{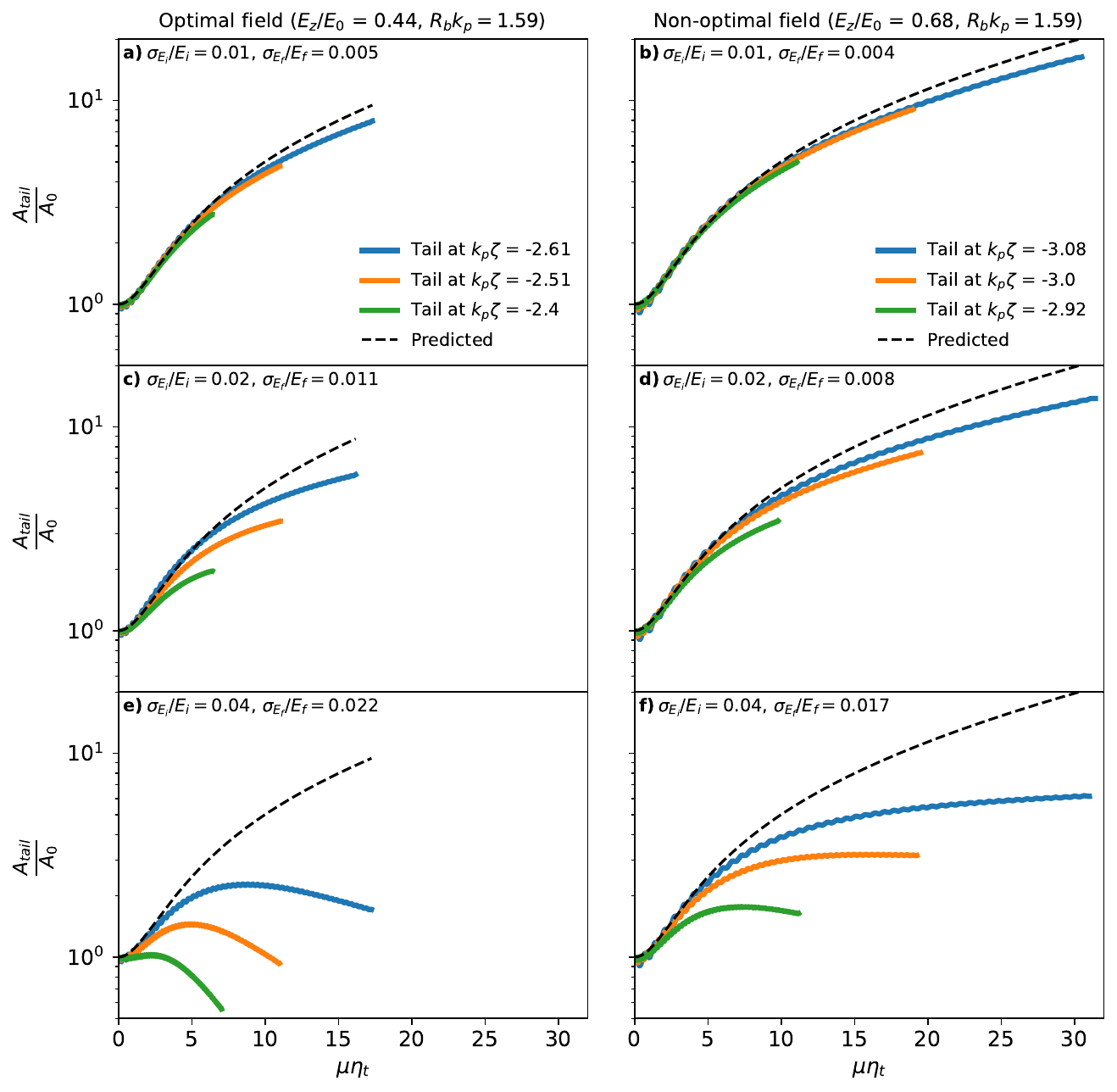}
    \caption{
        Oscillation-amplitude growth of the trailing bunch with optimal (left) and less-optimal (right) operating points at an efficiency of 0.4. (a, b) Oscillation-amplitude growth is shown for the trailing bunch with 1\% initial uncorrelated energy spread, (c, d) 2\% initial uncorrelated energy spread, and (e, f) 4\% initial uncorrelated energy spread. The initial and final energy spread in each case is shown in the top left corner panel. For the optimal operating point, increasing the initial energy spread leads to a reduction in the oscillation-amplitude growth until it reaches values that are less than at the initial time of acceleration. For the less-optimal point, the reduction is less pronounced. However, for the medium and short bunch lengths, we see that the oscillation-amplitude growth begins to plateau when the energy spread is 4\% (f).}
    \label{energyspreadamp}
\end{figure*}


\subsection{Uncorrelated energy spread}
Initial uncorrelated energy spread in the driving bunch has been shown to reduce the oscillation-amplitude growth from the hose instability~\cite{energyspreadhosing} due to phase-mixing of the electron betatron oscillations. We simulate the same configurations as in Fig.~\ref{fig:combined}, but with increasingly larger energy spread in the trailing bunch to investigate the effect on the BBU instability. This is shown in Fig.~\ref{energyspreadamp}.

We can see that initially, with 1\% energy spread, the oscillation-amplitude growth is reduced compared to Fig.~\ref{fig:combined} with no initial energy spread. For the same accelerating field $E_z$, the shorter bunches experience a greater reduction in their oscillation-amplitude growth such that they begin to show a large deviation from the expected growth. This is because the phase mixing of the betatron oscillation kicks in before significant oscillation-amplitude growth has had time to develop; the tails of the longer bunches experience much stronger wakefields, which causes a much faster growth rate in the oscillation-amplitude. The same behavior can be observed for the bunches with 2\% and 4\% energy spread but at a larger scale. Additionally for the optimal-$\eta_{t}$ case at an energy spread of 4\%, the phase mixing happens at a sufficiently small timescale compared to the instability such that the oscillation-amplitude growth of the bunches is damped to below 1 before the instability grows.

\begin{figure*}[t]
    \includegraphics[width=1\textwidth]{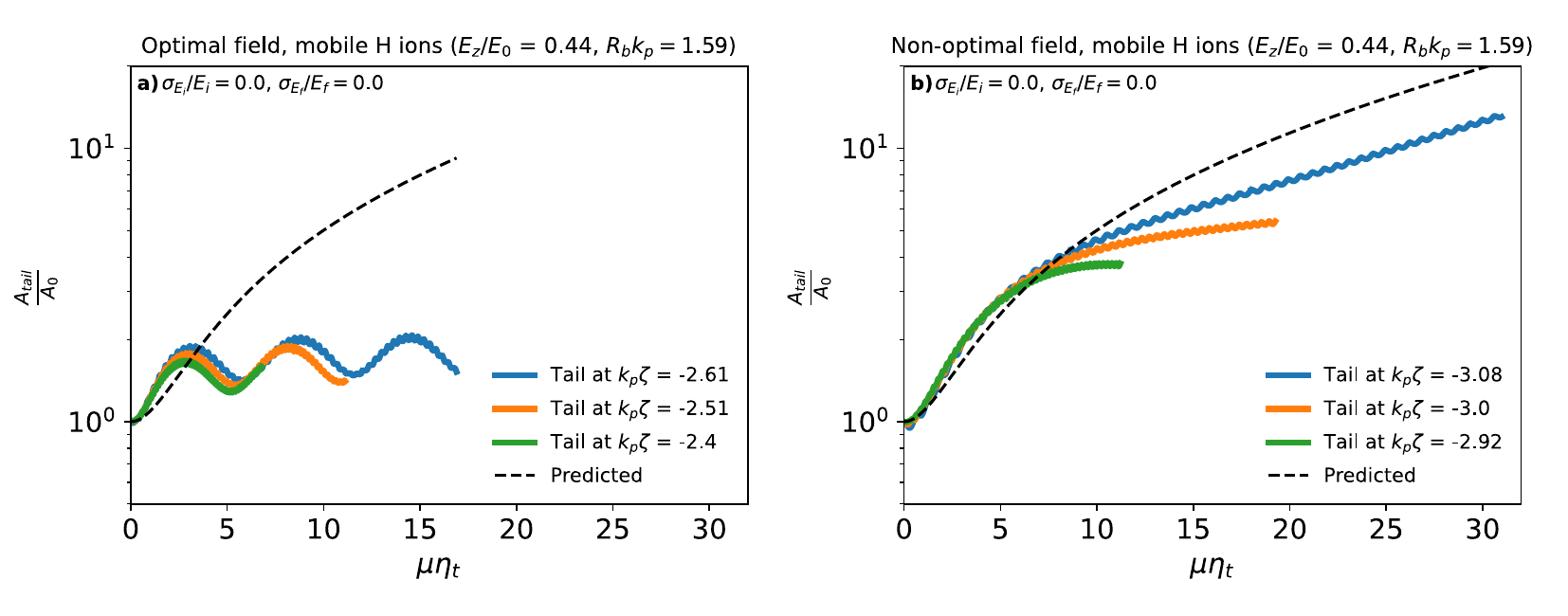}
    \caption{Tail oscillation-amplitude growth comparison between the case with optimal (a) and non-optimal field value (b) at $R_{b}k_{p} = 1.59$ with ion motion of hydrogen ions included. The simulation setup and parameters are identical to those in Fig.~\ref{fig:combined}, besides including ion motion. The optimal case with $E_{z}/E_{0} = 0.44$ (a) experiences significant damping of the oscillation-amplitude growth. The non-optimal case with $E_{z}/E_{0} = 0.68$ (b) also shows a reduced amplitude growth.}
    \label{fig:combined2}
\end{figure*}

In Fig.~\ref{energyspreadamp}(d), with 2\% energy spread, the oscillation-amplitude growth is reduced. However, it is still significant, reaching oscillation amplitudes approximately one order of magnitude larger than initially when considering the entire bunch. In Fig.~\ref{energyspreadamp}(f) with 4\% energy spread, the oscillation-amplitude growth is significantly reduced compared to the analytical growth when considering the entire bunch. For the medium and short bunch lengths the oscillation-amplitude growth begins to plateau at the end of the propagation distance.

As mentioned in the introduction, future collider applications typically require sub-percent energy spread after acceleration. This means that only the beams in Figs.~\ref{energyspreadamp}(a--d) are close to being viable for a collider application. However, this simulation only considers a single plasma stage. For a collider with several plasma stages, there are additional effects that would have to be taken into account, such as longitudinal movement of the electrons between stages \cite{longcorr}. Additionally, only the energy spread at the end of the acceleration stages is strictly constrained. Therefore, one could possibly have a large energy spread in the beam when going through the early stages of the accelerator, where the oscillation amplitude grows the fastest. Finally, as discussed in the next section, in collider applications the energy spread may be combined with ion motion to increase the damping.

\subsection{Ion motion}

If the electron-beam density in a plasma-based accelerator is large enough, the space-charge fields of the beam induce non-negligible motion of the background ions in the plasma~\cite{Ionmotion}. Recently, it has been shown that ion motion can mitigate the hose instability~\cite{ionsuppress}. Ion motion induces a head-to-tail variation in the focusing force, which induces phase mixing of the betatron oscillations of the electron bunch. Hence, the hose instability or the BBU instability can be suppressed by phase mixing breaking up the coherence of the oscillations. The optimal-$\eta_{t}$ and high-$\eta_{t}$ cases are now compared, with ion motion assuming hydrogen ions included. In these simulations, the transverse fields now contain a component induced by the ions, which increases the measured values of $\eta_{t}/\eta_{t, \mathrm{predicted}}$ to 4.5 and 4.6 (from 1.6 and 3.1, respectively). In order to better compare to the no-ion-motion simulations (Fig.~\ref{fig:combined}), we continue to employ the no-ion-motion values of $\eta_{t}$ when calculating $\mu\eta_{t}$ in Fig.~\ref{fig:combined2}. The figure shows that the amplitude growth is reduced compared to the predicted growth for most bunch lengths in both cases. As is the case without ion motion, the case with optimal field value generally sees a smaller amplitude growth than the non-optimal case. Here, the amplitude growth quickly reaches and subsequently oscillates around a plateau---a complex behavior caused by the nonlinear focusing fields of the ion-density spike. In the non-optimal case, the decoherence effect of ion motion is not large enough to counteract the rapid growth of the instability, but the amplitude growth does eventually reduce compared to no ion motion. Choosing the optimal accelerating gradient is therefore essential even in cases where ion motion suppresses the BBU instability. For ion motion to be viable as a damping mechanism in collider applications, one must prove that the instability is significantly damped and that emittance is preserved.

\section{Conclusions}

The efficiency--instability relation poses a significant challenge for high-efficiency acceleration of high-quality electron bunches. We find that in the ideal case with an optimally loaded (uniform) accelerating field, as well as no energy spread or ion motion, the previously proposed efficiency--instability relation in Eq.~\ref{etatetap} does not in fact represent either universal relation or an upper limit, but instead a lower limit on the strength of the instability for a given power-transfer efficiency. In short, we update the efficiency--instability relation to instead be an inequality given by
\begin{equation}
    \eta_{t} \geq \frac{\eta_{p}^{2}}{4(1-\eta_{p})},
    \label{newequality}
\end{equation}
which holds assuming no energy spread or ion motion.

A multidimensional parameter scan of PIC simulations was performed to find the optimal accelerating field that, for a given blowout radius, minimizes the transverse instability. At this field, which can be estimated with Eq.~\ref{optimalEz_low} to an accuracy of 14\% or better, the efficiency--instability relation in Eq.~\ref{etatetap} holds to within a small numerical factor.
However, if a non-ideal combination of blowout radius and accelerating field is chosen, the instability can be several orders of magnitude larger than that predicted by Eq.~\ref{optimalEz_low}. This can cause rapid oscillation-amplitude growth, which will lead to emittance growth and charge loss if not mitigated. Finally, the inclusion of sufficient initial energy spread in the trailing bunch, or employing a gas species which will give sufficient ion motion to mitigate the instability, can greatly reduce the oscillation-amplitude growth.

The initial energy spread needed for significant damping of the instability was shown to be larger than what is typically allowed in collider applications. However, this is for no ion motion in a single plasma stage. In a plasma accelerator used for collider applications one would need several plasma stages, for which the dynamics of the instability is non-trivial. Additional effects would have to be considered, such as longitudinal movement of the beam electrons between stages. Furthermore, one could combine the initial energy spread with additional damping mechanisms such as ion motion. A natural continuation of the work presented in this article would be to compute a similar parametric mapping, but focused on damping the amplitude growth for both single and multiple plasma stages.

Ultimately, the general strategy to minimize transverse instability should therefore first be to operate at the accelerating field given by Eq.~\ref{optimalEz_low}, and then to make use of either initial energy spread or non-negligible ion motion for additional suppression.

\begin{acknowledgements}
The authors thank B.~Foster for useful comments. This work was supported by the Research Council of Norway (NFR Grant No.~313770) and the European Research Council (ERC Grant No.~101116161). We
acknowledge Sigma2 - the National Infrastructure for High-Performance Computing and Data
Storage in Norway for awarding this project access to the LUMI supercomputer, owned by the
EuroHPC Joint Undertaking, hosted by CSC (Finland) and the LUMI consortium.
\end{acknowledgements}

\end{document}